# Authentication and identity management based on zero trust security model in micro-cloud environment

Ivana Kovacevic[0000-0001-9418-8814], Milan Stojkov[1][0000-0002-0602-0606], and Milos Simic[1][0000-0001-8646-1569]

[1] Faculty of Technical Sciences, Novi Sad 21000, Republic of Serbia
kovacevic.ivana@uns.ac.rs

**Abstract.** The abilities of traditional perimeter-based security architectures are rapidly decreasing as more enterprise assets are moved toward the cloud environment. From a security viewpoint, the Zero Trust framework can better track and block external attackers while limiting security breaches resulting from insider attacks in the cloud paradigm. Furthermore, Zero Trust can better accomplish access privileges for users and devices across cloud environments to enable the secure sharing of resources. Moreover, the concept of zero trust architecture in cloud computing requires the integration of complex practices on multiple layers of system architecture, as well as a combination of a variety of existing technologies. This paper focuses on authentication mechanisms, calculation of trust score, and generation of policies in order to establish required access control to resources. The main objective is to incorporate an unbiased trust score as a part of policy expressions while preserving the configurability and adaptiveness of parameters of interest. Finally, the proof-of-concept is demonstrated on a micro-cloud platform solution.

## 1 Introduction

In the digital era, IoT and mobile devices interconnect using heterogeneous protocols in both human-centric and machine-centric networks. Taking into consideration the significant number of connected devices and massive data traffic, it became inevitable to incorporate cloud services in many domains, from smart cities and self-driving vehicles to social networks and the gaming industry. Nevertheless, the increased amount of raw data being generated in real-time created a bottleneck in meeting the required quality of services due to these devices' computational, storage, and bandwidth constraints. Therefore, there is a strive to move data centers to the edge, closer to data resources, in order to decrease latency and increase on-demand network access, introducing the concept of edge computing. The basic idea of edge computing is to employ a hierarchy of edge servers with increasing computation capabilities to handle mobile and heterogeneous computation tasks offloaded by the low-end IoT and mobile devices, namely, edge devices [1]. Such advantages over cloud computing have caused edge computing to grow rapidly in recent years. As the general interest and number of users of cloud platforms increase, the security of cloud services is slowly becoming a requirement of uttermost importance.

Along with data migration, new consideration was born. Since customers have a high demand for privacy and restricted access control over their assets, many security challenges have arisen. In this context, there is a need to develop a flexible security service that could be deployed with other cloud services to provide them with mandatory security features. The abilities of traditional perimeter-based security architectures are rapidly decreasing as more enterprise assets are moved toward the cloud environment. From a security viewpoint, the zero trust framework can better track and block external attackers while limiting security breaches resulting from insider attacks in the cloud paradigm. Furthermore, Zero Trust can better accomplish access privileges for users and devices across cloud environments to enable the secure sharing of resources.

Multiple trust domains coexist in the cloud computing environment, and numerous user entities participate in communication and interaction. Therefore, implementing authentication of the application system is crucial [3]. User authentication in cloud computing is the process of validating the identity of the user to ensure that it is legitimate to access cloud resources [4]. Authentication is a critical aspect of security given that it protects from the impersonification of entities, preventing attackers from illegally accessing the system. Authentication in cloud computing (PCC) is classified into six types: username and password authentication (password-based authentication), multifactor authentication, mobile trusted, Single Sign On, Public Key Infrastructure, and biometric authentication. Despite the growing number of innovative ways to authenticate users, password-based authentication is still one of the most popular methods [5]. Besides, from the aforementioned, authentication is generally performed between devices and servers. If an attacker intends to access protected servers or devices directly, it would be blocked by the authentication system.

In this paper, we proposed a novel micro-service Zero Trust architecture based on multiple technologies to enhance the security of the system. Our proposed architecture incorporates a multi-factor authentication mechanism for authenticating users, as well as certificate management performed by PKI component. Furthermore, we employ an attribute-based authorization. The key aspect of our approach is the utilization of a trust score in conjunction with access rules to facilitate access control. By incorporating a trust score as an additional factor, we enhance the decision-making process for granting access.

The remainder of the paper is organized as follows: Sect. 2 provides a brief overview of the related work considering zero trust architecture. In Sect. 3, the proposed zero trust model is explained, emphasizing the services for user and device authentication, the definition of policies using the ABAC model, and the trust score calculation. Subsequently, in Sect. 4, the conclusion is presented along with the future directions for research.

## 2      Related work

Many solutions have been put forth with the objective of effectively integrating or mitigating the cloud environment's security in order to meet the requirements of a zero-trust architecture. Syed et al. [6] gave a detailed overview of the Zero Trust (ZT) security paradigm. The article employs a descriptive approach to present the core principles of ZTA, predominantly focusing on the role of authentication and access control. The authors contributed an in-depth discussion of cutting-edge authentication and access control techniques in diverse scenarios. This paper impacted our work by explaining various methods for calculating the level of trust, along with an overview of relevant markup languages used for defining access control policies in attribute-based access control models.

Another concise overview of the various measures that need to be taken into account when implementing Zero-Trust Architecture (ZTA) is given in [7]. The paper elaborates on the essential steps required to achieve continuous authentication and authorization for all network participants. Additionally, the authors introduce a trust engine component that dynamically calculates the overall trust of a user, device, or application within a specific network, assigning it a trust score.

Dimitrakos et al. [8] introduced a new model for trust-aware continuous authorization and a novel technology implementing this model that is scalable, efficient, and lightweight enough to be effective in a consumer IoT setting. Their implementation focuses on trust in authentication and follows a trust fusion approach where information from many sources is combined. Such an approach is formalized in Subjective Logic (SL). Generally, in cases with multiple sources of belief, it can be assumed that a collection of different beliefs can reflect the ground truth better than each belief independently. A formal description of multi-source fusion is presented in [9] and can be referenced for a formal description when deciding on parameter weights in the proposed trust algorithm in this paper.

## 3      Zero trust security model

Zero Trust encompasses a set of concepts aimed at reducing or ideally eliminating the ambiguity associated with enforcing precise access decisions for each request by considering the network as compromised. Zero Trust Architecture (ZTA), in turn, refers to the specific system design intended to facilitate this objective. ZTA involves the implementation of a wide range of fundamental principles to secure enterprise assets, including data, devices, users, and infrastructure components. The key principles for achieving ZTA are authentication and access control, as these are the means by which the user's identity is established [6]. ZTA is based on the concepts of least privilege, granular access control, and dynamic and strict policy enforcement, ensuring that no user or device is implicitly trusted, regardless of status or location.

The Zero Trust model is based on continuous or context-based authentication. Communication is secure regardless of network location, meaning all messages transferred between nodes are encrypted. Popular authentication methods include symmetric key authentication, lightweight public key infrastructure (PKI), and Open Authorization 2.0 [6]. In asymmetric key authentication, digital certificates can be utilized to prove the identity of a device before communication is established.

### 3.1    Proposed zero trust architecture

Our approach is based on multiple technologies, including multi-factor authentication, public key infrastructure, and attribute-based access control (ABAC) expressed by XACML [10] policy language. The proposed model consists of several interconnected components responsible for implementing different security requirements. The responsibility of the authentication service is to handle requests from both users and devices. Its implementation is designed to support various protocols, specifically HTTPS and gRPC secure, ensuring secure communication. NIST suggests using software-defined perimeters as one of the key strategies in ensuring the effective implementation of ZTA [11]. The authentication service represents the entry point into the software-defined perimeter (SDP) zone, where each request is intercepted and verified.

All relevant information from requests is collected and transferred to the logging component, which is responsible for integration with the security information and event management (SIEM) system. The segregation of logging activities into a distinct service is undertaken with the objective of ensuring enhanced scalability and performance within the system architecture. By separating the logging functionality, the primary components are freed from the responsibility of performing logging operations, thereby enabling them to concentrate on their core features. To achieve loose coupling and mitigate potential blocking or delayed responses, the communication between the logging service and other components within the system is designed to be asynchronous through the utilization of a message queue. By implementing asynchronous communication through a message queue, the logging service can operate independently and interact with other components in a decoupled manner. That enables IAM services to perform tasks without causing delays or impacting the overall system's responsiveness.

Upon successful authentication, entities are redirected to the policy enforcement point, a separate service that aggregates relevant logs regarding the entities' historical behavior. These logs are then prepared and made available to the policy engine (PE). The policy engine consists of functions for defining and managing access policies, using domain-specific language for writing policies. Trust score is calculated in a separate function within the PE, based on aggregated inputs from PEP. Subsequently, the trust score is compared against an empirically established predefined threshold. Policy engine, authorization entry point (AEP), authentication service, and PKI service are all part of the IAM system, which can be described as the framework of policies and technologies that ensure authorized people within an organization have the appropriate access to network resources [14]. That is a micro-service-oriented architecture utilizing in-line context enrichment and publish-subscribe protocols to optimize concurrency among various components such as policy parsing, attribute value retrieval, trust level determination, and policy evaluation.

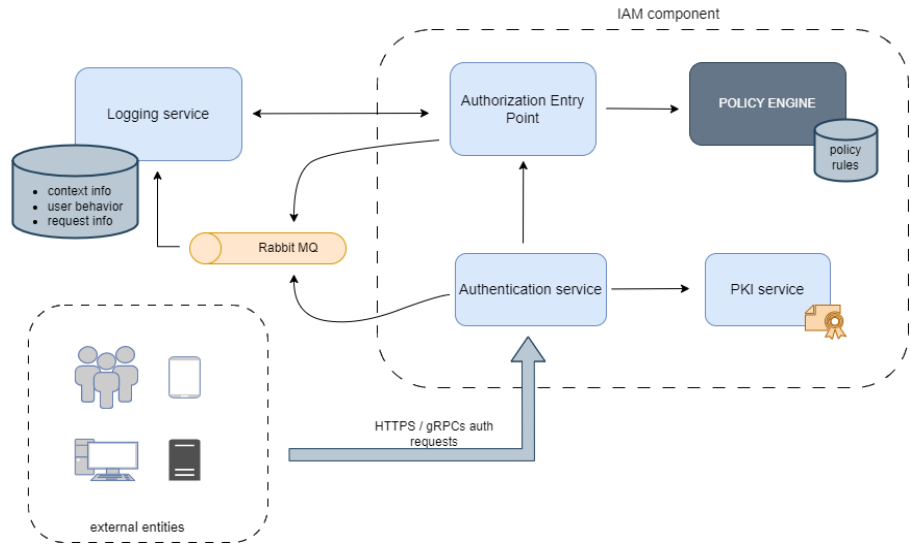

**Fig. 1.** Proposed zero trust architecture

## 3.2 Authentication of users

Authentication enables the various entities participating in a distributed environment to confirm each other's identities before initiating actual communication. For user authentication, a multi-factor authentication scheme is proposed. The trust in authenticity increases exponentially when more factors are involved in the verification process [12]. The initial stage of user authentication verifies credentials, namely username and password. However, additional information apart from credentials validation is necessary during the authentication process. The authentication service is responsible for the extraction of contextual details such as the user's geographical location, the IP address associated with the request, and the timestamp. This context-specific information is then transmitted to a logging service and stored in a database. It can serve as a factor within an implicit trust mechanism, working in conjunction with other parameters like the time duration since the last successful login.

Relying solely on contextual information for implicit authentication is not advisable for a cloud environment, given the diversity of the users, devices, and resources involved. Rather than introducing expensive or vendor-specific solutions, we have chosen to utilize one-time software tokens. As a second factor, a soft token is generated through an authenticator application. This approach does not require deployment complexity and comes with a low cost making it appropriate for cloud environments.

### 3.3    Authentication of devices

Node authentication and verification are implemented through the Zero Trust security model. Nodes provide certificates upon joining the cloud computing environment. Trusted certificates are stored centrally and managed by PKI service as a part of the IAM component. The IAM component is developed as an identity-based certificate management system. It enables securing, storing, and controlling secrets, passwords, certificates, and encryption keys for protecting users and sensitive data. PKI service also implements mechanisms for signing and verifying digital signatures. Sensitive information, including encryption keys and secrets, is stored relying on Vault [13].

### 3.4    Definition of Attribute-based access control policies

The ABAC can be defined as an access control model where subjects' requests to perform operations on objects are evaluated by considering the attributes of the subject, object, and environment, along with the policies defined around these attributes and conditions. Extensible Access Control Markup Language (XACML) supports and implements the ABAC model. XACML defines policies by using logical formulae involving attributes. The attribute concept in ABAC offers a flexible and extensible abstraction for capturing characteristics of things and keeping access rules abstract enough to be applicable to heterogeneous resources [8].

XACML comes with a robust architecture consisting of multiple software components responsible for creating and managing policies, rules, advice, and obligations. Policy administration point (PAP) creates a policy or set of policies, and it is designed to manage the lifecycle of a policy. The policy decision point (PDP) component is responsible for taking authorization requests as input and returning one of the following responses: PERMIT, DENY, or INDETERMINATE. It is important to note that access decision functions are fed from two different sources. Some user data is stored within the Policy Information Point (PIP). The PIP serves as a repository for granted or denied permits, making it accessible for policy evaluation and decision-making when needed. This separation allows efficient and controlled access to user information within the XACML architecture. To prevent the unnecessary transfer of a large volume of contextual data with each authorization request in XACML, the PIP service is periodically updated with aggregated user data collected in the logging service. PIP also stores records of unsuccessful access attempts that resulted in denial, as well as records of initially granted access to a resource. The latter information is essential, especially because in the proposed ZTM, distinct policy rules are applied based on whether users have an existing authorization history or are new to the system.

Our solution employs a hybrid authorization model, combining criteria and a score-based trust algorithm (TA) along with applying policy enforcement. The criteria-based algorithm is used on users without relevant behavior history, requiring all factors to be fulfilled before evaluating policy rules. This approach enforces strict policies specifically for newly registered users or those requesting access to a particular resource for the first time. In contrast, the score-based trust algorithm calculates a score by assigning weights to various attributes and compares this score against a predetermined threshold value. By continuously successfully passing the authorization mechanism, the user can be awarded for his behavior by transiting to score-based evaluation. The authorization is performed in cycles, solely determined by users' past behavior and current location within a predefined period. Subsequently, the entire process is reinitiated, ensuring a continuous evaluation and adaptation based on the user's behavior history. An example of a condition for enabling an organization member to READ resources if he is in a perimeter of 100km from the resource from the same organization is shown in Figure 2.

```
<Condition>
    <Apply FunctionId=
"urn:oasis:names:tc:xacml:3.0:function:string-equal">
     <Apply FunctionId=
"urn:oasis:names:tc:xacml:3.0:function:map-distance">
     <Apply FunctionId=
"urn:oasis:names:tc:xacml:1.0:function:double-bag">
       <Apply FunctionId=
  "urn:oasis:names:tc:xacml:1.0:function:string-bag">
        <AtributeValue>
           {user_lat},{user_long}
        </AttributeValue>
       </Apply>
       <Apply FunctionId=
  "urn:oasis:names:tc:xacml:1.0:function:string-bag">
        <AttributeValue>
           {resource_lat},{resource_long}
        </AttributeValue>
       </Apply>
     </Apply>
     </Apply>
     <AttributeValue>100</AttributeValue>
    </Apply>
</Condition>
```

**Fig. 2.** Access rule condition example

### 3.5    Calculation of trust score

Trust assessment plays a crucial role in trust-based access control for cloud computing. Nonetheless, a significant challenge lies in assigning reasonable and unbiased weights to various trust factors or parameters involved in the assessment process. The algorithm for calculating trust can be thought of as a function with inputs being relevant attributes such as access request type, the previous behavior of the requesting entity, resource usage history, prior history of penalties, current IAM policies, trust of the group to which the entity belongs to and present threat scenario [15]. Our approach considers the following factors as TA inputs: request metadata (time of request, current details of service or application being used to make the request, and IP address), number of previous requests made to the same resource, resource usage history, previous history of penalties and request geo-location. Aggregated data comes from AEP to the policy engine for each authorization request. The final decision is made considering the result from XACML PDP and the calculated trust score compared against the specified threshold. If PDP returns the PERMIT value and the trust score passes the threshold boundary, access to the resource is granted. Table 1 shows the possible outcomes of the evaluation.

**Table 1.** Possible outcomes from trust evaluation

|  | Trust score >= threshold | Trust score < threshold |
|---|---|---|
| PERMIT | allowed | denied |
| DENY | denied | denied |
| INDETERMINATE | re-evaluate rules | denied |

## 4    Discussion

This section discusses and compares all techniques mentioned above with other possible solutions. Multifactor authentication is an effective method for establishing a robust authentication mechanism without compromising the user experience or relying on dedicated hardware components to safeguard sensitive information. It eliminates the need for distributing and managing physical tokens, resulting in potential cost savings. On the other side, certificates provide a strong security mechanism for device authentication. They utilize public-key cryptography to establish trust between the device and the authentication server. Since all computational work is transferred and performed in a cloud environment, there is no need to use lightweight encryption algorithms. Although the initial setup required effort, certificates with external key storage using a vault scaled effectively on a micro-cloud platform. The use of certificates simplifies the management of device identities, and the vault handles the storage and retrieval of keys efficiently, enabling secure authentication for a large number of devices. To enable the granularity of access policy rules, it is argued that more than role-based access control is needed. In that sense, RBAC can be seen as a special case of ABAC. XACML involves multiple complex features for condition matching, resolution of conflicts, and computation of rules, which enables a strong base for extending and adapting policies in the future. Separating the trust score from access rules allows greater flexibility in defining and adjusting authorization decisions. A separate trust score also enables adaptive authorization by allowing the system to dynamically adjust access privileges based on the current trust level of the user or entity. Given that we already gather and log various contextual information and event histories, it is possible to utilize this data as input for a machine learning (ML) algorithm as a part of future work. By leveraging

ML techniques, we can fine-tune the trust assessment process and improve the accuracy of the trust score calculation.

## 5 Conclusion

The implementation of the proposed zero trust approach involves a distinct IAM component comprising multiple services. This component is integrated with an existing micro-cloud platform for evaluation. For user authentication, MFA is suggested as an additional layer of security. In contrast, device authentication is accomplished through a two-way SSL authentication process, wherein digital certificates are provided and verified. The system incorporates fine-grained attribute-based access policies that are integrated with the calculated trust score to achieve adaptiveness and continuous verification. The authorization process operates continuously, adhering to the principle of "never trust - always verify" as prescribed by the NIST report on ZTA [11]. Even though the proposed proof of concept promises essential safeguarding of a cloud computing environment, there are certain improvements to be considered, namely regarding performance and coverage of other aspects zero-trust maturity model (ZTMM), including intrusion detection systems, continuous event monitoring, and network segmentation.

**Acknowledgment.** Funded by the European Union (TaRDIS, 101093006). Views and opinions expressed are however those of the author(s) only and do not necessarily reflect those of the European Union. Neither the European Union nor the granting authority can be held responsible for them.

## References

1. Xiao, Yinhao, et al. "Edge computing security: State of the art and challenges." Proceedings of the IEEE 107.8 (2019): 1608-1631.
2. Furfaro, Angelo, Alfredo Garro, and Andrea Tundis. "Towards security as a service (secaas): On the modeling of security services for cloud computing." 2014 international carnahan conference on security technology (ICCST). IEEE, 2014.
3. LI, Xinghua, et al. Smart applications in edge computing: Overview on authentication and data security. IEEE Internet of Things Journal, 2020, 8.6: 4063-4080.
4. Eldow, Abdalla, et al. "Literature review of authentication layer for public cloud computing: a meta-analysis." Journal of Theoretical and Applied Information Technology 97 (2006): 12.
5. Shen, Chao, et al. "User practice in password security: An empirical study of real-life passwords in the wild." Computers & Security 61 (2016): 130-141.
6. Syed, Naeem Firdous, et al. "Zero trust architecture (zta): A comprehensive survey." IEEE Access (2022).
7. S. Mehraj and M. T. Banday, "Establishing a Zero Trust Strategy in Cloud Computing Environment," 2020 International Conference on Computer Communication and Informatics (ICCCI), Coimbatore, India, 2020, pp. 1-6, doi: 10.1109/ICCCI48352.2020.9104214.

8. Dimitrakos, Theo, et al. "Trust aware continuous authorization for zero trust in consumer internet of things." 2020 IEEE 19th international conference on trust, security and privacy in computing and communications (TrustCom). IEEE, 2020.

9. Wang, Dongxia, and Jie Zhang. "Multi-source fusion in subjective logic." 2017 20th International Conference on Information Fusion (Fusion). IEEE, 2017.

10. OASIS XACML Homepage, https://www.oasis-open.org/committees/tc_home.php?wg_abbrev=xacml, last accessed 2023/05/23

11. Rose, Scott, et al. Zero trust architecture. No. NIST Special Publication (SP) 800-207. National Institute of Standards and Technology, 2020.

12. Babaeizadeh, Mahnoush, Majid Bakhtiari, and Alwuhayd Muteb Mohammed. "Authentication methods in cloud computing: A survey." Research Journal of Applied Sciences, Engineering and Technology 9.8 (2015): 655-664.

13. Vault Homepage, https://www.vaultproject.io/, last accessed 2023/05/23.

14. Carnley, P. Renee, and Houssain Kettani. "Identity and access management for the internet of things." International Journal of Future Computer and Communication 8.4 (2019): 129-133.

15. Sarkar, Sirshak, et al. "Security of zero trust networks in cloud computing: A comparative review." Sustainability 14.18 (2022): 11213.